\documentclass[12pt]{article}
\usepackage[utf8]{inputenc}
\usepackage{amsmath,amssymb}
\usepackage{booktabs}
\usepackage{algorithm}
\usepackage{graphicx}
\usepackage{algorithmic}
\usepackage[font={small,it}]{caption}
\usepackage{geometry}
\usepackage{hyperref}
\usepackage{amsthm}
\usepackage{float}
\hypersetup{breaklinks=true}

\title{A Comprehensive Corpus of Biomechanically Constrained Piano Chords: Generation, Analysis, and Implications for Voicing and Psychoacoustics}

\author{Mahesh Ramani\\
\textit{Independent}}

\date{January 2026}

\begin{document}

\maketitle

\begin{abstract}
I present the generation and analysis of the largest known open-source corpus of playable piano chords (approximately 19.3 million entries). This dataset enumerates the two-handed search space subject to biomechanical constraints (two hands, each with 1.5 octave reach) to an unprecedented extent. To demonstrate the corpus's utility, the relationship between voicing shape and psychoacoustic targets was modeled. Harmonicity proved intrinsic to pitch-class identity: voicing statistics added negligible variance ($\Delta R^2 \approx 0.014\%$, $p \approx 0.13$). Conversely, voicing significantly predicted dissonance ($\Delta R^2 \approx 6.75\%$, $p \approx 0.0008$). Crucially, skewness ($\beta \approx +0.145$) was approximately 5.8$\times$ more effective than spread ($\beta \approx -0.025$) at predicting roughness. The analysis challenges the pedagogical emphasis on ``spread'': skewness is a stronger predictor of dissonance than spread. This suggests that clarity in ``open voicings'' is driven less by width than by negative skewness; achieving lower-register clearance by placing wide gaps at the bottom and allowing tighter clustering in the treble. The results demonstrate the corpus's ability to enable future research, especially in areas such as generative modeling, voice-leading topology, and psychoacoustic analysis.
\end{abstract}

\textbf{Keywords:} computational musicology, piano chords, voicing statistics, psychoacoustics, dissonance, harmonicity

\section{Introduction}

Data-driven approaches to musicology and music generation rely heavily on the quality and scale of underlying corpora \cite{adebayo2023}. While symbolic music datasets have grown in size, there remains a gap between theoretical chord spaces and instrument-specific chord spaces, which are bound by the physical limitations of the performer. The distinction is critical for the piano: a chord that is theoretically consonant is musically irrelevant if it cannot be executed by human hands. Consequently, the lack of large-scale, biomechanically constrained datasets has limited the ability of researchers to model how the physical arrangement of notes (voicing) influences psychoacoustic perception in a realistic performance context.

To address this scarcity, this paper presents the generation and analysis of the largest known open-source corpus of playable piano chords. Unlike datasets derived from transcription, which are biased by stylistic prevalence, or random MIDI sampling, which ignores physical constraints, this corpus exhaustively enumerates the two-handed search space spanning MIDI 21--108. The generation process enforces biomechanical hand-span limits (1.5 octaves per hand, a generous upper bound) and note-count constraints, ensuring that every entry represents a physically viable sonic event.

To demonstrate the analytical utility of this dataset, I present a validation study modeling the relationship between the statistical ``shape'' of a voicing and its psychoacoustic properties. Specifically, I investigate a common pedagogical heuristic: the principle of ``open voicing,'' which suggests that spreading notes across a wider range reduces perceptual roughness. By calculating two acoustic metrics -- Harmonicity (as in, closeness to the harmonic series) and Plomp-Levelt Dissonance \cite{plomp1965} (extrinsic sensory roughness) -- I evaluate whether statistical moments of voicing (centroid, spread, skewness, and kurtosis) offer predictive power beyond the pitch-class identity of the note.

The results of this validation serve two purposes. First, they confirm that the corpus captures subtle acoustic variances necessary for machine learning tasks through providing a quantitative benchmark. Second, the analysis offers a novel music-theoretic insight: while voicing statistics are negligible predictors of harmonicity, they significantly predict dissonance. Furthermore, the data suggests that the pedagogical emphasis on chord ``spread'' (width) is mathematically less precise than chord ``skewness'' (asymmetry) in predicting consonance. Specifically, pianists should aim for negative skewness to reduce dissonance -- placing wide gaps in the lower register while allowing tighter clustering in the treble.

These findings establish the corpus as a robust resource for future research in chord generation, graph topology, and psychoacoustic modeling, which is further discussed in Section 4.

\section{Methodology}

\subsection{Dataset Generation and Search Space}

I defined the search space using a piano-inspired model to the following constraints:

\begin{enumerate}
\item \textbf{Range:} The pitch space is bounded by the standard 88-key piano range (MIDI 21--108, $A_0$ to $C_8$).
\item \textbf{Hand Span:} To ensure playability, each hand is modeled as a subset of notes fitting within a sliding window of 19 semitones (approximately a 1.5-octave span, or a major tenth).
\item \textbf{Polyphony:} A chord is defined as the set union of notes generated by a Left Hand (LH) and a Right Hand (RH), where each hand contributes at least one note.
\end{enumerate}

Hands may overlap in pitch range without restriction, as any unrealistic fingering (e.g., crossed hands in dense clusters) can trivially be refingered equivalently by reassigning notes from left to right while preserving the sonic event.

\subsubsection{Sampling Strategy}

Due to the combinatorial explosion of possible note combinations as chord density increases, I employed a hybrid generation strategy:

\begin{itemize}
\item \textbf{Exhaustive Enumeration ($N_{\text{notes}} \in [1, 5]$):} For chords containing 1 to 5 notes, I exhaustively enumerated every valid combination of two-handed positions. This ensures the dataset contains the complete universe of playable low-density chords.
\item \textbf{Monte Carlo Sampling ($N_{\text{notes}} \in [6, 10]$):} For chords containing 6 to 10 notes, the theoretical search space exceeds computational feasibility. I therefore employed Monte Carlo sampling (Seed=42) to generate 1,000,000 unique, valid instances for each cardinality.
\end{itemize}

The resulting dataset, released as manus-piano-chord-corpus \cite{ramani2026data}, contains $\sim$19,376,000 unique voicings, balanced across the density spectrum.

\subsection{Feature Extraction}

For every generated chord, I computed two classes of features: statistical descriptors of the voicing (shape) and psychoacoustic target metrics.

\subsubsection{Voicing Statistics (Input Features)}

I treated the MIDI note numbers as a distribution and calculated the statistical moments to quantify the ``shape'' of the voicing:

\begin{itemize}
\item \textbf{Centroid:} The arithmetic mean of the MIDI pitches.
\item \textbf{Spread:} The range (max minus min) in semitones.
\item \textbf{Skewness:} The asymmetry of the note distribution (3rd standardized moment). Positive skew indicates clustering in the bass with outliers in the treble, and vice versa.
\item \textbf{Kurtosis:} The ``tailedness'' of the distribution (4th standardized moment), measuring the tendency toward clusters versus uniform spacing.
\end{itemize}

These moments are formally defined as follows for a set of MIDI note values $m_1, m_2, \dots, m_n$ (where $n \geq 2$) with centroid $\mu$ and standard deviation $\sigma$:

\begin{align}
\text{Skewness} &= \frac{1}{n} \sum_{i=1}^n \left( \frac{m_i - \mu}{\sigma} \right)^3 \\
\text{Kurtosis} &= \left[ \frac{1}{n} \sum_{i=1}^n \left( \frac{m_i - \mu}{\sigma} \right)^4 \right] - 3
\end{align}

For single-note or degenerate cases ($n < 2$ or $\sigma \approx 0$), skewness defaults to 0 and kurtosis to $-3$ (a sentinel value indicating no meaningful distribution). It should be noted that the constant component of the kurtosis equation serves as a correction for the baseline of a normal distribution, and kurtosis $= -3$ isn't achieved in practice.

To ensure these voicing statistics were independent of the pitch-class and note-count controls, they were residualized -- that is, I regressed each voicing feature on the control variables and extracted the residuals (the unexplained portion). This procedure removes any variance in voicing shape that is merely a byproduct of which notes are played or how many notes are present, allowing us to isolate the unique contribution of spatial arrangement.

I also computed the Interval Class (IC) Vector \cite{forte1973} for every chord to serve as a control variable, ensuring that analysis could separate the effects of pitch-class identity from vertical voicing. The IC Vector is a 12-element array representing the distribution of interval classes (modulo 12 semitones) among all unique pairs of notes in a chord. Each element at index $i$ (0 to 11) counts the number of intervals equivalent to $i$ semitones modulo an octave. The fixed length of 12 arises from 12-TET structure, where intervals larger than a tritone (6 semitones) are equivalent to their inversions (e.g., 7 semitones $\equiv$ 5 downward).

\subsubsection{Psychoacoustic Metrics (Target Variables)}

\paragraph{1. Plomp-Levelt Dissonance (Roughness):}

I implemented the Plomp-Levelt model (1965) \cite{plomp1965} to estimate sensory dissonance based on the interference of partials. I modeled pairwise dissonance $d$ between two frequencies $f_1, f_2$ (where $f_1 < f_2$) using the standard curve parameters:

\begin{equation}
d(x) = e^{-3.5x} - e^{-5.75x}
\end{equation}

where $x$ is the frequency difference scaled by the critical bandwidth ($x = s \cdot (f_2-f_1)$). $s$, the scaling factor, is defined as 
\begin{equation} 
    \frac{0.24} {(0.0207 \cdot \min(f_1,f_2) + 18.96)} 
\end{equation}. To ensure computational efficiency across 19 million entries, an $88 \times 88$ interaction matrix was precomputed for all piano keys. The total dissonance of a chord was calculated as the sum of all pairwise interactions.

Frequencies derive from MIDI notes using 12-TET tuning with A4 = 440 Hz. This ignores piano-specific inharmonicity and assumes additivity of pairwise interactions, as validated in the original perceptual studies.

\paragraph{2. Harmonicity:}

Harmonicity is computed as a measure of how closely a chord's frequencies align with a single harmonic series, using a best-fit fundamental frequency ($f_0$) search. For each candidate $f_0$ (derived as $f_i / n$ for note frequency $f_i$ and integer $n=1$ to 12), the mean fractional error is calculated as:

\begin{equation}
\text{mean\_err} = \frac{1}{k} \sum_{j=1}^k |q_j - \text{round}(q_j)|
\end{equation}

where $q_j = f_j / f_0$ and $k$ is the number of notes. The minimum mean\_err across candidates is selected; ``closely align'' is implicitly defined with a tolerance of $10^{-6}$ (if mean\_err $< 10^{-6}$, harmonicity = 1.0 for near-perfect matches, accounting for floating-point precision).

The score is normalized as:

\begin{equation}
\text{Harmonicity} = 1 - 2  \min(\max(\text{mean\_err}, 0), 0.5)
\end{equation}

mapping the theoretical error range [0, 0.5] to [1, 0] for intuitive interpretation (1 = perfect harmonic fit, 0 = maximally inharmonic).

This approach is similar to template-matching models in psychoacoustics, such as Parncutt's (1988) revision of Terhardt's virtual pitch model, which evaluates chord roots via harmonic pattern matching with fractional deviations from integer multiples \cite{parncutt1994}.

\subsection{Limitations}

While the corpus provides a robust foundation for computational musicology, several limitations should be noted: (1) Equal loudness assumption across notes, ignoring dynamic variations or register-dependent amplitude in real performance; (2) Modeling of pure sine waves without overtones, neglecting piano string inharmonicity that could alter dissonance and harmonicity in practice; (3) Lack of perceptual validation through human listener studies, relying instead on established psychoacoustic models that may not fully capture subjective consonance; (4) Biomechanical constraints based on a generous 1.5-octave span, potentially including edge cases less playable for average hand sizes; (5) No consideration of temporal or contextual factors, such as chord duration or surrounding harmony, which influence real-world perception. Future work could address these via empirical listening tests or timbre-inclusive extensions.

\section{Results}

\subsection{The Intrinsic Nature of Harmonicity}

I tested whether statistical descriptors of voicing (centroid, spread, skewness, and kurtosis) predict a chord's harmonicity after controlling for its pitch-class composition. If harmonicity is purely a function of the intervals present (e.g., a ``Major 7th'' chord), voicing shape should provide little to no predictive power.

By employing a baseline regression model controlling for pitch-class content (represented by IC Vector) and note count, I tested whether adding voicing features improved predictions on held-out chords. The analysis yielded a strong null result regarding the influence of voicing:

\begin{itemize}
\item Pitch-class content alone explains the overwhelming majority of predictable variance ($R^2 \approx 0.77$).
\item Adding the block of residualized voicing moments yielded a vanishingly small increase in out-of-sample explained variance ($\Delta R^2 \approx 0.00014$). In percentage terms, voicing statistics explain only an additional 0.014\% of the variance.
\item A non-parametric permutation test confirmed that this contribution is not statistically significant ($p \approx 0.13$).
\end{itemize}

The effect sizes for individual moments were practically negligible. For instance, a one-standard-deviation increase in residualized kurtosis or spread corresponds to a change in harmonicity of approximately $+0.001$ to $+0.002$ on a 0--1 scale.

\begin{figure}[H]
\centering
\fbox{\includegraphics[width=0.6\textwidth]{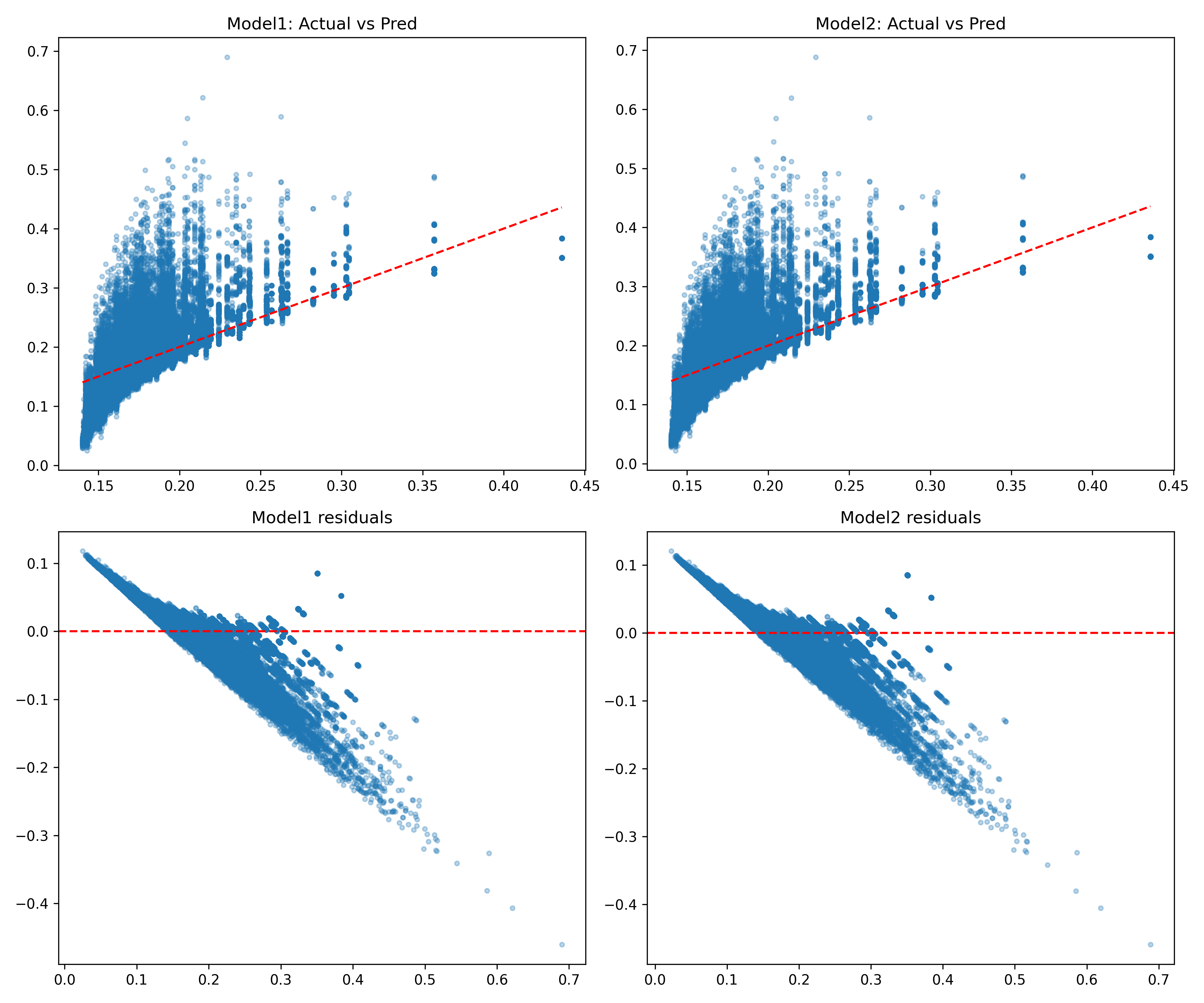}}
\caption{Residual plot for harmonicity predictions (Model 2). Residuals show no systematic patterns as a function of predicted values, confirming model adequacy. Importantly, the residual distribution remains virtually identical whether voicing moments are included (Model 2) or excluded (Model 1), consistent with the negligible $\Delta R^2 \approx 0.00014$.}
\label{fig:harmonicity}
\end{figure}

This confirms that harmonic fit is intrinsic to pitch-class identity. Unlike sensory dissonance, which varies based on physical note spacing, harmonicity is robust to voicing transformations. A chord's spectral lock is determined almost entirely by what notes are played, not how they are distributed across the keyboard.

\subsection{Dissonance}

As done with harmonicity, I modeled the Plomp-Levelt dissonance of a chord as a function of its statistical moments (centroid, spread, skewness, kurtosis), controlling for its specific pitch-class identity. However, to isolate the effect of voicing ``shape'' from pitch-class ``content,'' I first trained a baseline model using only the Interval Class Vector and note count. I then introduced the four spatial moments (centroid, spread, skewness, kurtosis) as additional features. This approach allows me to answer the question: does how notes are arranged add predictive power beyond what notes are played?

\begin{figure}[H]
\centering
\fbox{\includegraphics[width=0.6\textwidth]{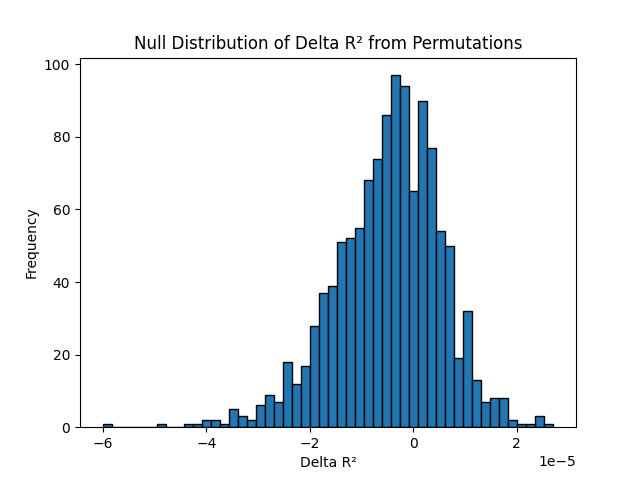}}
\caption{Permutation test for the contribution of voicing moments to dissonance prediction. The histogram shows the null distribution of $\Delta R^2$ values obtained by randomly permuting the residualized voicing features (1,200 iterations). The observed $\Delta R^2 = 0.0675$ falls far in the right tail ($p \approx 0.0008$), strongly rejecting the null hypothesis that voicing shape does not contribute to dissonance beyond pitch-class content.}
\label{fig:permutation}
\end{figure}

While pitch-class content explains the majority of the variance ($R^2 \approx 0.642$), the addition of voicing moments produced a robust out-of-sample improvement in fit, raising the explained variance to $R^2 \approx 0.710$ ($\Delta R^2 = 6.75\%$). Permutation testing yielded a significance level of $p \approx 0.0008$, strongly rejecting the null hypothesis as shown in Figure \ref{fig:permutation}. Unlike harmonicity (where voicing was negligible), the spatial topology of a chord is a statistically significant driver of sensory dissonance. Put simply: the same pitch classes arranged differently on the keyboard will produce measurably different levels of roughness.

The key finding emerges from the regression coefficients:

\begin{itemize}
\item \textbf{Spread:} Standardized $\beta \approx -0.025$ (statistically significant, but practically negligible).
\item \textbf{Skewness:} Standardized $\beta \approx +0.145$ (a robust predictor of dissonance).
\end{itemize}

Teachers and textbooks widely advocate for ``spreading'' notes (increasing the range/width of a chord) to reduce muddiness. However, it is shown that spread is a mathematically poor proxy for consonance. To put these coefficients in practical terms: positive skewness (clustering notes in the bass) increases dissonance. Conversely, negative skewness reduces dissonance approximately 5.8 times more effectively (derived from the absolute value of the division of the beta values) than increasing spread by one standard deviation.

These results indicate that the perceived clarity of an ``open voicing'' is not driven by using the whole keyboard (Spread), but rather by achieving negative skewness in the note distribution. A negatively-skewed voicing implies large gaps in the lower register and tighter clustering in the treble.

In summary, what teachers call ``spread'' is better understood as asymmetric spacing that follows the Overtone Series principle: wide intervals at the bottom, narrow intervals at the top. While sheer width offers a marginal reduction in dissonance, negative skewness is more effective at predicting the reduction of psychoacoustic roughness. It could be said that this finding reframes a common teaching heuristic in more precise, measurable terms.

\begin{figure}[H]
\centering
\fbox{\includegraphics[width=0.6\textwidth]{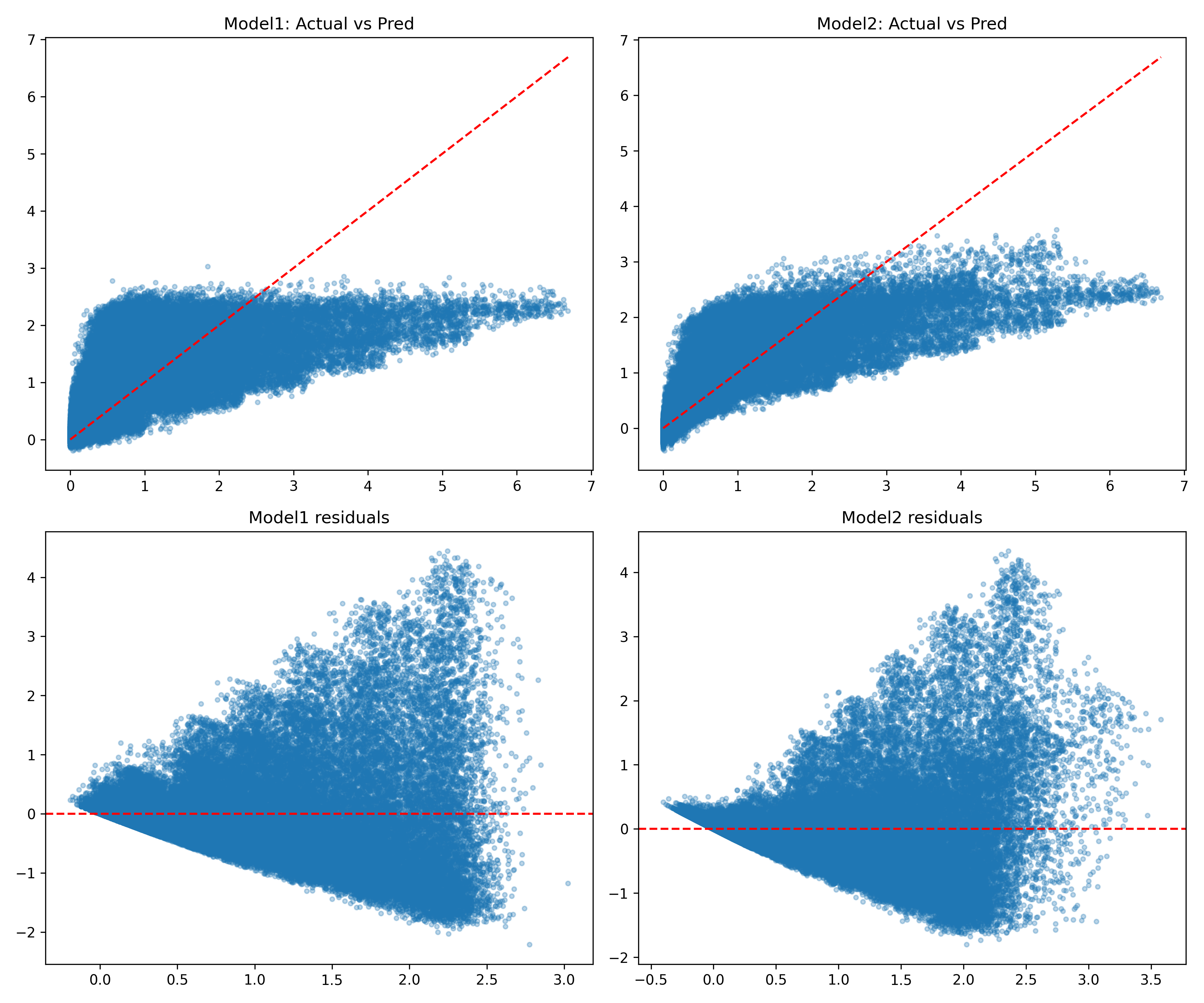}}
\caption{Actual vs. predicted dissonance values for Model 2 (controls + voicing moments) on the held-out test set. Each point represents one chord (N = 10,000 test chords). The diagonal line represents perfect prediction. The model achieves $R^2 \approx 0.71$, demonstrating that voicing shape, when combined with pitch-class content, provides substantial predictive power for psychoacoustic roughness.}
\label{fig:dissonance}
\end{figure}

\section{Discussion}

\subsection{Intrinsic vs. Extrinsic Musical Properties}

It can be established from Section 3 that harmonicity is an ``intrinsic'' musical property, while dissonance is an extrinsic one.

My null result for Harmonicity demonstrates that the mathematical fit of a chord is hard-coded into its pitch-class identity -- that is, no amount of voicing manipulation can fundamentally alter the spectral ratio of a C-Major triad, for example. Conversely, the positive result for Dissonance ($\Delta R^2 \approx 6.75\%$) proves that sensory roughness is highly malleable.

The distinction has important implications for computational models. Harmony analyzers, for example, can safely ignore voicing when estimating chord roots or tonal function, while generative systems must explicitly model spatial arrangement to control perceptual roughness. This validates the separation of ``harmony'' (the abstract selection of notes) and ``voicing'' (the physical realization of those notes) as distinct cognitive tasks, where the former dictates function and the latter dictates texture.

\subsection{Refining Pedagogical Intuitions}

As shown in 3.2, the pervasive instruction to ``open up the voicing'' is revealed here to be a case of semantic imprecision. While the advice successfully leads students to reduce dissonance, the regression coefficients suggest that spread is not the mechanism driving this improvement.

So, as opposed to this common instruction, the data offers more precise advice: clear the lower register of small intervals while allowing notes to cluster more tightly in the treble. This aligns with established psychoacoustic theory regarding critical bandwidths, which are significantly wider in the bass frequencies, necessitating larger spatial gaps to avoid roughness.

\subsection{Applications and Future Directions}

These findings directly demonstrate the necessity of large-scale, instrument-specific datasets. Previous studies on voicing have often been limited by small sample sizes where variables like Spread and Skew are highly collinear, making it difficult to disentangle their effects. By exhaustively enumerating playable chords as described 2.1, this corpus breaks that collinearity, allowing for the rigorous isolation of voicing features.

Beyond theoretical analysis, the corpus offers immediate utility for algorithmic arrangement and performance modeling. Current generative models often output MIDI data that ignores human biomechanics; this dataset provides a comprehensive `dictionary' of valid hand shapes (fingerings, for example, could potentially be reconstructed from looking at the IC Vector), allowing systems to quantize abstract harmonic output into physically playable piano reductions. Furthermore, the granular dissonance scoring enables narrative tension profiling -- that is, it could allow the analysis of historical works by the specific psychoacoustic `roughness curves' created by the composer's voicing choices as opposed to harmonic progression.

\subsubsection{Graph Topology and Voice Leading}

One particularly promising direction involves network analysis of the chord space itself. The corpus provides the necessary nodes to construct a comprehensive ``voice leading graph,'' where connections are defined by voice-leading distance or quality of transition. Preliminary graph-theoretic analysis performed on a subset of 20,000 samples suggests that the playable piano space exhibits strong ``Small-World'' network properties, yielding a Small-World Coefficient ($\sigma$) of approximately 7 \cite{watts1998}. This high coefficient validates the intuition that efficient voice leading allows a pianist to navigate between disparate harmonic regions with minimal physical displacement. Future work can leverage larger sample sizes to map the ``highways'' of idiomatic piano performance. This would empirically verify which chord transitions are easiest and identify difficult outliers.

\subsubsection{Machine Learning and Generative Systems}

The dataset serves as suitable content for ML/AI purposes. By mapping abstract Interval Class vectors to specific, evaluated voicings, the corpus provides the labeled data necessary to train models that respect both biomechanical constraints and psychoacoustic targets. Generative systems can utilize the provided dissonance scores as objective functions (or reward signals), learning to predict voicings that maximize clarity for any given harmonic input.

\section{Conclusion}

This paper presented the generation and analysis of a comprehensive corpus of 19.3 million playable piano chords, demonstrating its utility for isolating psychoacoustic effects with high statistical power. The results establish a clean separation: harmonicity is intrinsic to pitch-class identity, while dissonance is extrinsic and highly sensitive to spatial arrangement. Most notably, the corpus's scale enabled the first rigorous decoupling of skewness from spread, revealing that asymmetric note distribution---not sheer width---drives the clarity of open voicings. This reframes a ubiquitous pedagogical heuristic in precise, measurable terms and validates the corpus's ability to support fine-grained psychoacoustic analysis.

This corpus is freely available for future research (hosted on Hugging Face \cite{ramani2026data}) with potential applications including voice-leading graph analysis, training generative models that respect biomechanical constraints, and empirical validation studies with human listeners.


\begin{thebibliography}{9}

\bibitem{plomp1965}
Plomp, R., \& Levelt, W. J. M. (1965). Tonal consonance and critical bandwidth. \textit{The Journal of the Acoustical Society of America}, 38(4), 548--560. \url{https://doi.org/10.1121/1.1909741}

\bibitem{watts1998}
Watts, D. J., \& Strogatz, S. H. (1998). Collective dynamics of `small-world' networks. \textit{Nature}, 393(6684), 440--442. \url{https://doi.org/10.1038/30918}

\bibitem{forte1973}
Forte, A. (1973). \textit{The structure of atonal music}. Yale University Press.

\bibitem{barbancho2013}
Barbancho, A. M., Barbancho, I., Tardón, L. J., \& Barbancho, A. M. (2013). \textit{Database of piano chords: An engineering view of harmony}. Springer. \url{https://doi.org/10.1007/978-1-4614-7476-0}

\bibitem{adebayo2023}
Adebayo, T. S., \& Flierl, M. (2023). Jazznet: A dataset of fundamental piano patterns for music audio machine learning research. In \textit{ICASSP 2023 - 2023 IEEE International Conference on Acoustics, Speech and Signal Processing (ICASSP)} (pp. 1--5). IEEE. \url{https://doi.org/10.1109/ICASSP49357.2023.10096620}

\bibitem{napolitano2023}
Napolitano, N., Orsenigo, D., Cannas, I., Bailes, F., Didonna, V., Mauri, G., \& Pachet, F. (2023). ChoCo: A chord corpus and a data transformation workflow for musical harmony knowledge graphs. \textit{Scientific Data}, 10(1), 635. \url{https://doi.org/10.1038/s41597-023-02410-w}

\bibitem{kantarelis2024}
Kantarelis, S., \& Nikolaidis, N. (2024). Chordonomicon: A dataset of 666000 chord progressions. \textit{arXiv preprint arXiv:2410.22046}.

\bibitem{parncutt1994}
Parncutt, R., \& Strasburger, H. (1994). Applying psychoacoustics in composition: ``Harmonic'' progressions of ``nonharmonic'' sonorities. \textit{Perspectives of New Music}, 32(2), 88--129.

\bibitem{sethares1993}
Sethares, W. A. (1993). Local consonance and the relationship between timbre and scale. \textit{Journal of the Acoustical Society of America}, 94(3), 1218--1228.

\bibitem{ramani2026data}
M. Ramani, ``manus-piano-chord-corpus,'' Hugging Face, 2026. [Dataset]. DOI: \url{https://doi.org/10.57967/hf/7506}

\end{thebibliography}
\end{document}